\documentclass[aps,pre,twocolumn,groupedaddress]{revtex4}
\usepackage{graphicx}
\usepackage{amsmath}
\usepackage{amssymb}

\begin{document}

\title{Fluctuation Theorems on Nishimori Line}
\author{Masayuki Ohzeki}
\affiliation{Department of Systems Science, Kyoto University, Yoshida-Honmachi, Sakyo-ku,
Kyoto 606-8501, Japan}
\date{\today }

\begin{abstract}
The distribution of the performed work for spin glasses with gauge symmetry is considered. 
With the aid of the gauge symmetry, which leads to the exact/rigorous results in spin glasses, we find a fascinating relation of the performed work as the fluctuation theorem. 
The integral form of the resultant relation reproduces the Jarzynski-type equation for spin glasses we have obtained. 
We show that similar relations can be established not only for the distribution of the performed work but also that of the free energy of spin glasses with gauge symmetry, which provides another interpretation of the phase transition in spin glasses.
\end{abstract}
\maketitle

\section{Introduction}
The fluctuation theorem makes current activities to understand the nonequilibrium behavior \cite{Evans1993,Evans1994,Gallavotti1995a,Gallavotti1995b,Kurchan1998,Lebowitz1999,Maes1999,Crooks1998}. 
The theorem consists of a relation between distribution functions in different conditions. 
The first discovery was in long time observation for an entropy production rate \cite{Evans1993,Evans1994}. 
The rigorous derivation of the fluctuation theorem was given by Gallavotti and Cohen for the thermostated deterministic steady-state ensembles \cite{Gallavotti1995a,Gallavotti1995b} and for the stochastic dynamics \cite{Kurchan1998,Lebowitz1999,Maes1999}. 
The fluctuation theorem for the nonequilibrium behavior as well as for the symmetry broken states in equilibrium have been discovered \cite{Gaspard2012}.
In the present study, we focus on the fluctuation theorem for the performed work distributions \cite{Crooks1998}.
This type of the fluctuation theorem yields a fascinating relation for the expectation of the exponentiated work known as the Jarzynski equality \cite{Jarzynski1997a,Jarzynski1997b}.
The Jarzynski equality makes a relationship between the equilibrium free energy differences associated with the initial and final conditions and the performed work during a nonequilibrium process.

Recently the author and co-worker investigated the Jarzynski equality for spin glasses with competing interactions between adjacent spins \cite{Ohzeki2010b,Ohzeki2011b}. 
Spin glasses often exhibit extremely long-time relaxation toward equilibrium. 
The long equilibration time hampers observations of the equilibrium state of spin glasses. 
However the previous study pointed out the possibility to investigate the equilibrium property from the observations in a different path through the nonequilibrium behavior.
Then we use the property of the Jarzynski equality by evaluating the average of the exponentiated performed work during the nonequilibrium process.
In other words, nonequilibrium behavior would not be nuisance but benefit to investigate the equilibrium behavior in spin glasses.
In the present study, we address the distribution of the performed work in such beneficial nonequilibrium behavior with long-time equilibration in spin glasses. 
Such fundamental research has never been performed as far as the best of our knowledge. 
We revisit the analysis of the nonequilibrium behavior in spin glasses in terms of the distribution function of the performed work by evaluation of the rate function. 
As a result, we obtain a relation in a similar form to the conventional fluctuation theorem for the performed work for spin glasses. 
The similar analyses reveal that the gauge symmetry, which leads to simple expressions of several quantities for spin glasses \cite{HNbook,Nishimori1981}, is closely related to the existence of the fluctuation-theorem type relation.

The paper is organized as follows.
The second section gives a brief introduction of several notations and tools in spin glasses.
In the third section, we review the previous study on the performed work by use of the Jarzynski equality.
In the present study we give the fluctuation theorem in spin glasses by employing several techniques developed in spin glasses.
We show the detailed analysis in \S IV.
The analyses with the aid of the specialized tool to spin glasses can yield other types of the fluctuation theorem rather than for the performed work.
We show the fluctuation theorems for the free energy differences and free energy itself in the following sections.
In the last section, we conclude our present work. 

\section{Spin glass and gauge symmetry}

We deal with the random-bond Ising model, whose Hamiltonian is defined as 
\begin{equation}
H(\mathbf{S}|\{\tau_{ij}\}) = - J\sum_{\langle ij\rangle
}\tau_{ij}S_{i}S_{j},  \label{Hami}
\end{equation}
where $S_{i}$ is the Ising spin taking values $\pm 1$, $J$ denotes strength of the coupling, and $\tau_{ij}$ is the sign of the coupling.
We use the notation $\mathbf{S}=(S_{1},S_{2},\cdots ,S_{N})$ for the spin configuration of total $N$ spins for convenience.
We set $J=1$ without loss of generality. 
The summation is taken over all bonds, whereas one may suppose usual nearest neighboring bonds on $d$-dimensional hyper-cubic lattice. 
We make no restrictions on the type or the dimension of the lattice in the present study. 
The distribution function of quenched randomness is specified as 
\begin{eqnarray}
P(\tau _{ij}) =p\delta (\tau _{ij}-1)+(1-p)\delta (\tau _{ij}+1) =\frac{%
\mathrm{e}^{\beta_{p}\tau _{ij}}}{2\cosh \beta_{p}},  \label{P2}
\end{eqnarray}
where $\beta_p$ is defined as $\mathrm{e}^{-2\beta_p}=(1-p)/p$. The
following analyses can readily be applied to other types of interaction and
its distribution functions as long as they satisfy a certain type of gauge
symmetry \cite{HNbook,Nishimori1981}.

We use the gauge transformation, which enables us to perform the exact/rigorous analyses in spin glasses especially on a special subspace $\beta=\beta_p$ known as the Nishimori line \cite{HNbook,Nishimori1981}. 
The gauge transformation is defined as 
\begin{eqnarray}
\tau_{ij} \to \tau_{ij}\sigma_i\sigma_j \quad S_i \to \sigma_iS_i,
\end{eqnarray}
where $\sigma_i$ is the gauge variable taking $\pm 1$. The Hamiltonian (\ref{Hami}) is gauge invariant, while the distribution function (\ref{P2}) changes as $P(\tau _{ij}) \propto \exp(\beta_p \tau_{ij} \sigma_i\sigma_j)$.
If we take the summation over all combinations of $\{\sigma_i\}$, the product of the distribution function for $\{\tau_{ij}\}$ over all bonds can be reduced to the partition function of the random-bond Ising model $Z(\beta_p;\{\tau_{ij}\})= \sum_{\sigma} \exp(\beta_p\sum_{\langle ij \rangle}\tau_{ij}\sigma_i\sigma_j)$. 
We mainly use this property to obtain the results in the present study.

In order to analyze the dynamical property of the Ising spin glass, let us suppose that the system evolves following a stochastic dynamics as governed by the master equation. 
In the present study, we consider to change the inverse temperature. 
This case is found in a generic solver of the optimization problem as the simulated annealing \cite{Kirkpatrick1983}. 
We will formulate our theory for discrete time steps for simplicity, although the continuous case can be treated similarly. We change the coupling $\beta$ from $\beta_0$ at $t=0$ to $\beta_T$ at $t=T$ in $T$ steps of time evolution, ($\beta_0, \beta_1, \cdots, \beta_T$). 
Correspondingly, the spin configuration changes as $\mathbf{S}_t$. 
The path probability for expressing the dynamical behavior of the system is then written by the product of the the transition rate from state $\mathbf{S}_{t}$ to state $\mathbf{S}_{t+1}$ following the master equation $P_t(\mathbf{S}_{t+1}|\mathbf{S}_t,\{\tau_{ij}\})$. 
The transition rate is also gauge-invariant, since it depends on the form of the Hamiltonian \cite{HNbook,Ozeki1995}.

\section{Jarzynski equality for spin glasses}
We analyze the work distribution function for spin glasses with a number of spins in the present study. 
For convenience, we fix several notations and review the previous study before demonstrating the detailed analysis on our issue.

\subsection{Jarzynski equality}
We define the discretized pseudo work given by the difference of the inverse temperature as 
\begin{equation}
\delta Y_t(\mathbf{S}_{t+1}|\{\tau_{ij}\}) = \beta_{t+1}H(\mathbf{S}_{t+1}|\{\tau_{ij}\}) - \beta_{t}H(\mathbf{S}_{t+1}|\{\tau_{ij}\}).
\end{equation}
If we change the parameters in the Hamiltonian instead of the inverse temperature, then the discretized pseudo work is reduced to the ordinary work as 
\begin{eqnarray}
\delta Y_t(\mathbf{S}_{t+1}|\{\tau_{ij}\}) &=& \beta H_{t+1}(\mathbf{S}_{t+1}|\{\tau_{ij}\},) - \beta H_t(\mathbf{S}_{t+1}|\{\tau_{ij}\}) \notag \\ 
&\equiv& \beta \delta W_t(\mathbf{S}_{t+1}|\{\tau_{ij}\}).
\end{eqnarray}
For instance, changing the magnetic field as in Ref. \cite{Ohzeki2011b} is the case. 
The performed (pseudo) work consists of collection of the discretized pseudo work 
\begin{equation}
Y(\{\mathbf{S}_{t}\},\{\tau_{ij}\}; 0 \to T) = \sum_{t=0}^{T-1} \delta Y_t(\{\tau_{ij}\},\mathbf{S}_{t+1}).
\end{equation}
Notice that the performed work depends on the specific configuration of $\{\tau_{ij}\}$.

We review the previous study for the performed work in spin glasses in short \cite{Ohzeki2010a,Ohzeki2011b}.
We applied the Jarzynski equality to the special case for spin glasses in the previous study.
The Jarzynski equality states that the expectation of the exponentiated work during nonequilibrium process is given by the difference of the free energy ($- \beta F = \log Z(\beta; \{\tau_{ij}\})$) between the initial and final conditions as \cite{Jarzynski1997a,Jarzynski1997b}
\begin{equation}
\langle {\rm e}^{-Y}\rangle_{0 \to T} = {\rm e}^{-\Delta_{0 \to T} (\beta F)},
\end{equation}
where $\Delta_{0 \to T} (\beta F) = \beta_TF(\beta_T;\{\tau_{ij}\})-\beta_0 F_0(\beta_0;\{\tau_{ij}\})$.
The brackets denote the nonequilibrium average over all realizations of the spin configurations in a nonequilibrium process starting from the equilibrium state defined as 
\begin{equation}
\langle \cdots \rangle_{0 \to T} = \sum_{\{\mathbf{S}_t\}} \prod_{t=0}^{T-1}
P_t(\mathbf{S}_{t+1}|\mathbf{S}_{t},\{\tau_{ij}\}) P^{t=0}_{\mathrm{eq.}}(\mathbf{S}_0|\{\tau_{ij}\}),
\end{equation}
where $P_{\mathrm{eq.}}^{t}(\mathbf{S}|\{\tau_{ij}\})$ denotes the equilibrium distribution function 
\begin{equation}
P_{\mathrm{eq.}}^{t}(\mathbf{S}|\{\tau_{ij}\}) = \frac{1}{Z(\beta_t;\{\tau_{ij}\})}\mathrm{e}^{ -\beta_t H(\mathbf{S}|\{\tau_{ij}\}) }.
\end{equation}

\subsection{For spin glasses}
Since the work is given by the realization of the spin configurations at each time, the path probability can be regarded as the distribution function of the performed work with the specific configuration of $\{\tau_{ij}\}$ as $P(y;\{\tau_{ij}\}, 0 \to T )$.
However, in spin glasses, we are often interested in the averaged quantity over all realizations of $\{\tau_{ij}\}$.
In the previous study, the author and co-worker evaluated the averaged Jarzynski equality as
\begin{equation}
\left[ \langle {\rm e}^{-Y}\rangle_{0 \to T}\right]_{\beta_0} = \left(\frac{2 \cosh \beta_T}{2 \cosh \beta_0}\right)^{N_B},\label{JENL0}
\end{equation}
where $N_B$ is the number of bonds, and the square brackets denote the configurational average over all realizations of $\{\tau_{ij}\}$ defined as
\begin{equation}
[ \cdots ]_{\beta_p} = \sum_{\{\tau_{ij}\}}\prod_{\langle ij \rangle} P(\tau_{ij}) \times \cdots = \sum_{\{\tau_{ij}\}}\prod_{\langle ij \rangle} \frac{%
\mathrm{e}^{\beta_{p}\tau _{ij}}}{2\cosh \beta_{p}} \times \cdots. 
\end{equation}
The above equality holds for the special initial condition that the nonequilibrium process starts from the Nishimori line $\beta_p = \beta_0$ \cite{Nishimori1981,HNbook}.
Then the averaged free energy on the right-hand side can be reduced to a trivial quantity.
However the Jarzynski equality is for the expectation, which is the average over all realizations.
We cannot obtain the detailed structure of the distribution functions only from the expectation.
In the present study, we thus revisit the problem on the performed work during nonequilibrium process in spin glasses by evaluating the distribution function in a different way.
That is the motivation of our study.

\section{Fluctuation theorem for spin glasses}
\subsection{Large deviation}
Throughout the present study, we assume the large deviation property in the distribution function for the system with a large number of components.
For instance, the distribution function $P(y;\{\tau_{ij}\}, 0 \to T )$ of the performed work for the specific configuration of $\{\tau_{ij}\}$ takes an asymptotic form as, for a large $N$, 
\begin{equation}
P(y;\{\tau_{ij}\}, 0 \to T ) \sim \mathrm{e}^{-N I(y;\{\tau_{ij}\}, 0 \to
T)},
\end{equation}
where $I(y;\{\tau_{ij}\}, 0 \to T)$ is the rate function and always takes a non-negative value. 
At the most frequent realization of the performed work (thermodynamic work), the rate function vanishes.
Here $y$ stands for the scaled work defined as 
\begin{equation}
y = Y(\{\mathbf{S}_{t}\},\{\tau_{ij}\}; 0 \to T)/N.
\end{equation}
In the thermodynamic system, the empirical average as the above scaled work can be evaluated by the zero point $y^*$ of the rate function and coincides with the expectation, since
\begin{equation}
\int dy P(y;\{\tau_{ij}\}, 0 \to T ) y = y^*.
\end{equation}
On the other hand, the rate function can characterize the fluctuation around the zero point.
For the rate function, an important relation, the fluctuation theorem, holds.
As detailed in Appendix \ref{AP1}, the fluctuation theorem states the symmetry of the rate functions as
\begin{eqnarray}
&&I(y; \{\tau_{ij}\}, 0 \to T)  \notag \\
&& = I(-y;\{\tau_{ij}\}, T \to 0)  \notag \\
&&\quad - y - \frac{1}{N} ( \log Z(\beta_T;\{\tau_{ij}\}) - \log
Z(\beta_0;\{\tau_{ij}\}) ).
\end{eqnarray}
This symmetry of the rate functions yields the well-known fluctuation theorem (Crook's fluctuation theorem) for the distribution functions of the performed work as 
\begin{equation}
\frac{P(y;\{\tau_{ij}\},0 \to T)}{P(-y; \{\tau_{ij}\}, T \to 0)}=\exp\left\{ N(y - \Delta_{0 \to T}(\beta f))\right\}.
\end{equation}
Rather than the Jarzynski equality, namely the expectation, the fluctuation theorem provides more detailed information on the distribution function.
Therefore we consider to find the fluctuation theorem to spin glasses in the present study.

\subsection{Generating function}
The above fluctuation theorem holds for the specific configuration of $\{\tau_{ij}\}$, similarly to the Jarzynski equality.
However, in spin glasses, we find sample-to-sample fluctuation in observations for different realizations of $\{\tau_{ij}\}$.
We thus must evaluate the fluctuation around the most probable realization for the infinite-size system. 
By the analysis of the generating function of the distribution function of $\{\tau_{ij}\}$ associated with the quantity we are interested in, we can evaluate such a sample-to-sample fluctuations by the rate function.
Let us define the following generating function of the performed work for spin glasses as 
\begin{equation}
\Omega_y(r;\beta_p,0 \to T) \equiv \frac{1}{N} \log \left[ \langle \exp(r N y ) \rangle_{0 \to T} \right]_{\beta_p},  \label{GSG}
\end{equation}
and of its inverse process, 
\begin{equation}
\Omega_y(r;\beta_p,T \to 0) \equiv \frac{1}{N} \log \left[ \langle \exp(-r N
y ) \rangle_{T \to 0} \right]_{\beta_p}.  \label{GSGInverse}
\end{equation}
It is convenient to define the generating function for the specific configuration of $\{\tau_{ij}\}$ as
\begin{equation}
\Psi_y(r;\{\tau_{ij}\},0 \to T) \equiv \frac{1}{N} \log \langle \exp(r N y )
\rangle_{0 \to T},  \label{GFW}
\end{equation}
and for its inverse process
\begin{equation}
\Psi_y(r;\{\tau_{ij}\},T \to 0) \equiv \frac{1}{N} \log \langle \exp(r N
(-y) ) \rangle_{T \to 0} .  \label{GFWInverse}
\end{equation}
Notice that the definition of $\Omega_y(r;\beta_p,0 \to T)$ reads, by Eq. (\ref{GFW}) 
\begin{eqnarray}
\exp(N\Omega_y(r;\beta_p,0 \to T)) &=& \left[ \langle \exp(r N y )
\rangle_{0 \to T} \right]_{\beta_p}  \notag \\
&=& \left[ \mathrm{e}^{N \Psi_y(r;\{\tau_{ij}\},0 \to T)} \right]_{\beta_p}.
\end{eqnarray}

For each realization of $\{\tau_{ij}\}$, the fluctuation theorem for the generating function holds as (See Appendix \ref{AP1})
\begin{eqnarray}
& & \Psi_y(r;\{\tau_{ij}\},0 \to T) = \Psi_y(-(r+1);\{\tau_{ij}\},T \to 0) 
\notag \\
& & + \frac{1}{N} \left(\log Z(\beta_T;\{\tau_{ij}\}) - \log
Z(\beta_0;\{\tau_{ij}\})\right).  \label{FT1}
\end{eqnarray}
We thus obtain 
\begin{eqnarray}
&&\mathrm{e}^{N \Psi_y(r;\{\tau_{ij}\}, 0 \to T)}  \notag \\
&&= \frac{Z(\beta_T;\{\tau_{ij}\})}{Z(\beta_0;\{\tau_{ij}\})}\mathrm{e}^{ N
\Psi_y(-(r+1);\{\tau_{ij}\},T \to 0)}.
\end{eqnarray}
Therefore we evaluate the exponentiated generating function as 
\begin{eqnarray}
&&\mathrm{e}^{N \Omega_y(r;\beta_p,0 \to T)}  \notag \\
&=& \left[\mathrm{e}^{N \Psi_y(r;\{\tau_{ij}\},0 \to T)}\right]_{\beta_p} 
\notag \\
&=& \left[\frac{Z(\beta_T;\{\tau_{ij}\})}{Z(\beta_0;\{\tau_{ij}\})} \mathrm{e%
}^{N \Psi_y(-(r+1);\{\tau_{ij}\}, T \to 0)}\right]_{\beta_p}\label{BGT1}.
\end{eqnarray}
The gauge transformation yields, as detailed in Appendix \ref{AP2}, 
\begin{eqnarray}
&& 2^N(2\cosh \beta_p)^{N_B}\mathrm{e}^{N \Omega_y(r;\beta_p,0 \to T)} 
\notag \\
&=& \sum_{\{\tau_{ij}\}}Z(\beta_p;\{\tau_{ij}\})\mathrm{e}^{N \Psi_y(r;\{\tau_{ij}\},0 \to
T)}  \notag \\
&=& \sum_{\{\tau_{ij}\}}\frac{Z(\beta_p;\{\tau_{ij}\})Z(\beta_T;\{\tau_{ij}%
\})}{Z(\beta_0;\{\tau_{ij}\})} \mathrm{e}^{ N \Psi_y(-(r+1);\{\tau_{ij}\},T \to 0)}.  \notag \\ \label{AGT1}
\end{eqnarray}
We set $\beta_p=\beta_0$ and then obtain 
\begin{eqnarray}
&&2^N(2\cosh \beta_0)^{N_B} \mathrm{e}^{N \Omega_y(r;\beta_0,0 \to T)} 
\notag \\
&=& \sum_{\{\tau_{ij}\}}Z(\beta_0;\{\tau_{ij}\})\mathrm{e}^{N \Psi_y(r;\{\tau_{ij}\},0 \to
T)}  \notag \\
&=& \sum_{\{\tau_{ij}\}}Z(\beta_T;\{\tau_{ij}\}) \mathrm{e}^{N
\Psi_y(-(r+1);\{\tau_{ij}\},T \to 0)}.\label{Eq1}
\end{eqnarray}
On the other hand, let us evaluate the exponentiated generating function of the inverse process 
\begin{eqnarray}
&&\mathrm{e}^{N \Omega_y(r;\beta_p,T \to 0)}  \notag \\
&=& \left[\mathrm{e}^{N \Psi_y(r;\{\tau_{ij}\},T \to 0)}\right]_{\beta_p} 
\notag \\
&=& \left[\frac{Z(\beta_0;\{\tau_{ij}\})}{Z(\beta_T;\{\tau_{ij}\})} \mathrm{e}^{N \Psi_y(-(r+1);\{\tau_{ij}\}, 0 \to T)}\right]_{\beta_p}.\label{BGT2}
\end{eqnarray}
When $\beta_p = \beta_T$, the gauge transformation yields
\begin{eqnarray}
& & 2^N(2\cosh \beta_T)^{N_B}\mathrm{e}^{N \Omega_y(r;\beta_T,T \to 0)} 
\notag \\
&=& \sum_{\{\tau_{ij}\}}Z(\beta_T;\{\tau_{ij}\})\mathrm{e}^{N \Psi_y(r;\{\tau_{ij}\},T \to
0)}  \notag \\
&=& \sum_{\{\tau_{ij}\}}Z(\beta_0;\{\tau_{ij}\}) \mathrm{e}^{N\Psi_y(-(r+1);\{\tau_{ij}\},0 \to
T)}.\label{Eq2}
\end{eqnarray}
Since the second line of Eq. (\ref{Eq1}) is equal to the third line of Eq. (\ref{Eq2}) except for the arguments of the generating function $\Psi_y$, we reach
\begin{eqnarray}  \label{JENL0}
\mathrm{e}^{N \Omega_y(-(r+1);\beta_0,0 \to T)} &=& \left(\frac{2\cosh
\beta_T}{2\cosh \beta_0}\right)^{N_B} \mathrm{e}^{N \Omega_y(r;\beta_T,T \to
0)}.  \notag \\ \label{FTg}
\end{eqnarray}
\subsection{Rate function and fluctuation theorem}
The relation (\ref{FTg}) yields the symmetry of the rate function of the performed work
for spin glasses.
We assume that the existence of the rate function for a large $N$ as
\begin{equation}
P(y;\beta_p,0 \to T) \sim \mathrm{e}^{-N J(y;\beta_p,0 \to T)}.
\end{equation}
From the generating function we evaluate the rate function through the Legendere transformation as 
\begin{equation}
J(y;\beta_p,0 \to T) = \sup_r \left\{ r y - \Omega_y(r;\beta_p,0 \to T)
\right\},
\end{equation}
and that for its inverse process as 
\begin{equation}
J(y;\beta_p,T \to 0) = \sup_r \left\{ r y - \Omega_y(r;\beta_p,T \to 0)
\right\}.
\end{equation}
Then the relation (\ref{JENL0}) yields 
\begin{eqnarray}
& & J(y;\beta_0,0 \to T)  \notag \\
&=& \sup_r \left\{ r y - \Omega_y(r;\beta_0,0 \to T) \right\}  \notag \\
&=& \sup_r \left\{ (r+1) y - \Omega_y(-(r+1);\beta_T,T \to 0) \right\} 
\notag \\
&& \quad -y - d \log\left(\frac{2\cosh \beta_T}{2\cosh \beta_0}\right) 
\notag \\
&=& J(-y;\beta_T,T \to 0) - y - d \log\left(\frac{2\cosh \beta_T}{2\cosh
\beta_0}\right).
\end{eqnarray}
where $d=N_B/N$. 
Consequently, we obtain the fluctuation theorem for spin glasses as 
\begin{equation}
\frac{P(y;\beta_0,0 \to T)}{P(y;\beta_T,T \to 0)} = \left(\frac{2\cosh
\beta_T}{2\cosh \beta_0}\right)^{N_B} \mathrm{e}^{N y}.  \label{FT2}
\end{equation}
The fluctuation theorem for spin glasses immediately reads 
\begin{equation}
\left[\left\langle \exp(-N y)\right\rangle_{0 \to T}\right]_{\beta_0} =
\left(\frac{2\cosh \beta_T}{2\cosh \beta_0}\right)^{N_B},  \label{JENL1}
\end{equation}
which reproduces the Jarzynski equality for spin glasses (\ref{JENL0}).
We here use the fact that we can replace the integration over all realizations of the performed work as the average over all configurations of $\{\tau_{ij}\}$ and $\{{\bf S}_{t}\}$ as
\begin{equation}
\left[ \langle \cdots \rangle_{0 \to T}\right]_{\beta_p} = \int dy P(y;\beta_p,0 \to T)\times \cdots.
\end{equation}

By taking logarithm of the fluctuation theorem (\ref{FT2}) and average to
make the form of the Kullback-Leibler (KL) divergence $D_{\mathrm{KL}%
}(P_A|P_B) = \int dx P_A(x)\log \left( P_A(x)/P_B(x)\right)$, we obtain two
inequalities as 
\begin{equation}
\left[ \langle y \rangle_{0 \to T} \right]_{\beta_0} + d \log\left( \frac{%
2\cosh \beta_T}{2\cosh \beta_0} \right) \ge 0.
\end{equation}
and 
\begin{equation}
\left[ \langle y \rangle_{T \to 0} \right]_{\beta_T} - d \log\left( \frac{%
2\cosh \beta_T}{2\cosh \beta_0} \right) \ge 0.
\end{equation}
Thus we obtain 
\begin{equation}
\left[ \langle y \rangle_{0 \to T} \right]_{\beta_0} + \left[ \langle y
\rangle_{T \to 0} \right]_{\beta_T} \ge 0 .  \label{SL1}
\end{equation}
The equality holds, when two of the distribution functions is the same, 
\begin{equation}
P(y;\beta_T,T \to 0)=P(y;\beta_0,0 \to T),
\end{equation}
since the left-hand side of Eq. (\ref{SL1}) can be evaluated by 
\begin{eqnarray}
&& \left[ \langle y \rangle_{0 \to T} \right]_{\beta_0} + \left[\langle y
\rangle_{T \to 0}\right]_{\beta_T}  \notag \\
&&= \int dy \left( P(y;\beta_0;0\to T) - P(-y;\beta_T,T \to 0)\right)  \notag
\\
& & \quad \times \log\left( \frac{P(y;\beta_0,0\to T)}{P(-y;\beta_T,T \to 0)}%
\right).  \label{KLy}
\end{eqnarray}
If we consider the quasi-static process, then $\langle y \rangle_{0 \to T} =
\Delta (\beta f)_{0 \to T}$ following the second law of thermodynamics (i.e. 
$P(y;\beta_p,0 \to T) = P(\Delta (\beta f);\beta_p,0 \to T)$). Therefore, for
the difference of the free energy of spin glasses, we
expect the existence of the similar relation to the above fluctuation theorem. 

\section{Fluctuation theorem for free energy difference}
\subsection{Generating function}
Let us consider the sample-to-sample fluctuations for the free energy
difference $\Delta (\beta f)$ by dealing with the rate function. We define
the generating function for the free energy difference as 
\begin{equation}
\Psi_{\Delta (\beta f)}(r;\beta_p,0 \to T) \equiv \frac{1}{N} \log \left[
\exp(r N \Delta_{0 \to T} (\beta f) ) \right]_{\beta_p}.
\end{equation}
The gauge transformation gives 
\begin{eqnarray}
&&\mathrm{e}^{N\Psi_{\Delta (\beta f)}(r;\beta_p,0 \to T)}  \notag \\
&&= \sum_{\{\tau_{ij}\}} \frac{Z(\beta_p;\{\tau_{ij}\})}{2^N(2\cosh
\beta_p)^{N_B}} \left(\frac{Z(\beta_T;\{\tau_{ij}\}) }{Z(\beta_0;\{\tau_{ij}%
\})}\right)^r.
\end{eqnarray}
We set $\beta_p = \beta_0$ and obtain 
\begin{eqnarray}
& & \mathrm{e}^{N\Psi_{\Delta (\beta f)}(r;\beta_0,0 \to T)}  \notag \\
& & = \sum_{\{\tau_{ij}\}} \frac{Z(\beta_T;\{\tau_{ij}\})}{2^N(2\cosh
\beta_0)^{N_B}} \left(\frac{Z(\beta_T;\{\tau_{ij}\}) }{Z(\beta_0;\{\tau_{ij}%
\})}\right)^{r-1}.
\end{eqnarray}
On the other hand the condition $\beta_p = \beta_T$ leads to 
\begin{eqnarray}
&& \mathrm{e}^{N\Psi_{\Delta (\beta f)}(r;\beta_T,0 \to T)}  \notag \\
&& = \sum_{\{\tau_{ij}\}} \frac{Z(\beta_T;\{\tau_{ij}\})}{2^N(2\cosh
\beta_T)^{N_B}} \left(\frac{Z(\beta_T;\{\tau_{ij}\}) }{Z(\beta_0;\{\tau_{ij}%
\})}\right)^r.
\end{eqnarray}
Therefore we find a relation of the generating function as 
\begin{eqnarray}
& & \Psi_{\Delta (\beta f)}(r+1;\beta_0,0 \to T)  \notag \\
& & = \Psi_{\Delta (\beta f)}(r;\beta_T,0 \to T) + d \log \left(\frac{2\cosh
\beta_T}{2\cosh \beta_0}\right).  \label{R1}
\end{eqnarray}
\subsection{Rate function and fluctuation theorem}
We assume that the large-deviation property holds for the free energy
of the large system $N \to \infty$ as 
\begin{equation}
P(\Delta_{0 \to T}(\beta f) ;\beta_p) \sim \exp\left( - N K(\Delta_{0 \to
T}(\beta f) ;\beta_p)\right).
\end{equation}
We then obtain the rate function of the free energy difference through the Legendre transformation as 
\begin{eqnarray}
& & K(\Delta_{0 \to T} (\beta f);\beta_p)  \notag \\
& & = \sup_t \left\{ r \Delta (\beta f) - \Psi_{\Delta (\beta
f)}(r;\beta_p,0 \to T) \right\}.
\end{eqnarray}
By use of Eq. (\ref{R1}), we immediately find 
\begin{eqnarray}
&&K(\Delta_{0 \to T} (\beta f);\beta_0)  \notag \\
&& = K(\Delta_{0 \to T}(\beta f);\beta_T) - \Delta( \beta f) - d \log \left(%
\frac{2\cosh \beta_T}{2\cosh \beta_0}\right).  \notag \\
\end{eqnarray}
Therefore we find 
\begin{equation}
\frac{P(\Delta_{0 \to T} (\beta f);\beta_0)}{P(\Delta_{0 \to T} (\beta
f);\beta_T)} = \left(\frac{2\cosh \beta_T}{2\cosh \beta_0}\right)^{N_B} 
\mathrm{e}^{N \Delta (\beta f)}.
\end{equation}
Due to the definition of the free energy difference, $P(\Delta_{0 \to T}
(\beta f);\beta_p) = P(-\Delta_{T \to 0} (\beta f);\beta_p)$, and we thus
obtain the relation in the same form as the fluctuation-theorem, 
\begin{equation}
\frac{P(\Delta_{0 \to T} (\beta f);\beta_0)}{P(-\Delta_{T \to 0} (\beta
f);\beta_T)} = \left(\frac{2\cosh \beta_T}{2\cosh \beta_0}\right)^{N_B} 
\mathrm{e}^{N \Delta (\beta f)}.  \label{FT3}
\end{equation}
We also establish the Jarzynski-type equality as, by integrating the
exponentiated free energy difference, 
\begin{equation}
\left[ \exp(-N \Delta_{0 \to T} (\beta f))\right]_{\beta_0} = \left(\frac{%
2\cosh \beta_T}{2\cosh \beta_0}\right)^{N_B}.
\end{equation}
We here use the fact that we can regard the distribution function of $\{\tau_{ij}\}$ as that of the free energy difference as
\begin{equation}
\left[ \cdots \right]_{\beta_p} = \int dx P(x:\beta_p)\times \cdots.
\end{equation}
By making the form of the KL divergence, we obtain the following
inequalities as 
\begin{equation}
\left[ \Delta_{0 \to T} (\beta f)\right]_{\beta_0} + d \log\left(\frac{%
2\cosh \beta_T}{2\cosh \beta_0}\right) \ge 0.
\end{equation}
and 
\begin{equation}
\left[ \Delta_{T \to 0} (\beta f)\right]_{\beta_T} - d \log\left(\frac{%
2\cosh \beta_T}{2\cosh \beta_0}\right) \ge 0.
\end{equation}
We thus obtain 
\begin{equation}
\left[ \Delta_{T \to 0} (\beta f)\right]_{\beta_T} + \left[\Delta_{0 \to T}
(\beta f)\right]_{\beta_0} \ge 0 .  \label{SL2}
\end{equation}
The deviation from zero can be written as 
\begin{eqnarray}
&& \left[ \Delta_{T \to 0} (\beta f)\right]_{\beta_T} + \left[\Delta_{0 \to
T} (\beta f)\right]_{\beta_0}  \notag \\
&&= \int dx \left( P(-x;\beta_T) - P(x;\beta_0)\right)\log\left( \frac{%
P(-x;\beta_T)}{P(x;\beta_0)}\right).  \notag \\
\end{eqnarray}
The equality holds when $P(\Delta_{0 \to T} (\beta f);\beta_0)=P(-\Delta_{T
\to 0} (\beta f);\beta_T)$. Since two of the distribution functions $%
P(\Delta_{0 \to T} (\beta f);\beta_0)$ and $P(-\Delta_{T \to 0} (\beta
f);\beta_T)$ do not coincide with each other in general, the equality in Eq.
(\ref{SL2}) is not expected to hold. Nor is the equality in Eq (\ref{SL1}).
The magnitude of the violation of the equality can be evaluated by the quantity related to the KL divergence.

As above considered the gauge symmetry leads to another type of the fluctuation theorem not only for the performed work but also the free energy differences for different configurations of $\{\tau_{ij}\}$. 
In this sense the obtained relations (\ref{FT2}) and (\ref{FT3}) are different from the ordinary fluctuation theorem. 
Our results are related to the sample-to-sample fluctuations of the different realizations of $\{\tau_{ij}\}$. 
The sample to sample fluctuation yields relevant effect even in equilibrium.
In theoretical studies in spin glasses, we usually employ the replica method to evaluate the equilibrium property.
In the next section we demonstrate to evaluate the equilibrium property for spin glasses with gauge symmetry from a perspective of the rate function without the replica method.
As a result, we find a different way to understand the peculiar behavior in spin glasses.

\section{Fluctuation theorem for free energy}
\subsection{free energy statistics}
We again assume that the large-deviation property for free energy in a large-$N$ system
\begin{equation}
P(f;\beta _{p},\beta )\sim \exp \left( -NL(f;\beta _{p},\beta )\right) .
\end{equation}%
Here we regard the distribution function of $\{\tau_{ij}\}$ as that of the free energy.
We define the generating function of the free energy as 
\begin{equation}
\Psi _{f}(r;\beta _{p},\beta )=\frac{1}{N}\log \left[ \exp (rN(-\beta f))%
\right] _{\beta _{p}}.
\end{equation}%
The exponentiated generating function is
\begin{equation}
\mathrm{e}^{N\Psi _{f}(r;\beta _{p},\beta )}=\left[ Z^{r}(\beta ;\{\tau
_{ij}\}))\right] _{\beta _{p}}.
\end{equation}%
The analysis by the gauge transformation lead us to 
\begin{eqnarray}
&&\Psi _{f}(r;\beta _{p},\beta )  \notag \\
&=&-\log 2-d\log \left( 2\cosh (\beta _{p})\right)   \notag \\
&&\quad +\frac{1}{N}\log \left( \sum_{\tau _{ij}}Z(\beta _{p};\{\tau
_{ij}\})Z^{r}(\beta ;\{\tau _{ij}\})\right) .
\end{eqnarray}%
On the Nishimori line $\beta _{p}=\beta $, we find 
\begin{eqnarray}
\Psi _{f}(r;\beta ,\beta ) &=&-\log 2-d\log \left( 2\cosh \beta \right)  
\notag \\
&&+\frac{1}{N}\log \left( \sum_{\tau _{ij}}Z^{r+1}(\beta ;\{\tau
_{ij}\})\right) .
\end{eqnarray}%
We obtain the following similar quantity in the symmetric distribution ($%
\beta _{p}=0$) 
\begin{equation*}
\Psi _{f}(r;0,\beta )=-d\log 2+\frac{1}{N}\log \left( \sum_{\tau
_{ij}}Z^{r}(\beta ;\{\tau _{ij}\})\right) .
\end{equation*}%

\subsection{Fluctuation theorem}
We find a relation from the above generating functions
\begin{equation}
\Psi _{f}(r;\beta ,\beta )+\log 2+d\log \left( \cosh \beta \right) =\Psi
_{f}(r+1;0,\beta ),
\end{equation}%
which is essentially the same as the calculation in Ref. \cite{Georges1987}.
The above relation enables us to analyze the critical behavior of the spin
glasses in the symmetry distribution through the free energy on the
Nishimori line as shown in Ref. \cite{Ohzeki2009b}. However, in a modern
point of view, this relation can be regarded as the fluctuation theorem for free energy. 
Indeed we find the symmetry of the rate function of
the free energy through the Legendre transformation as 
\begin{equation}
L(f;\beta ,\beta )=L(f;0,\beta )+\beta f+\log 2+d\log \left( \cosh \beta
\right),
\end{equation}%
where we defined the rate function as 
\begin{equation}
L(f;\beta _{p},\beta )=\sup_{r}\left\{ r(-\beta f)-\Psi _{f}(r;\beta
_{p},\beta )\right\} .  \label{Rate2}
\end{equation}%
Thus we obtain the fluctuation-theorem type equality for the free energies
on the Nishimori line and in the symmetric distribution as 
\begin{equation}
\frac{P(f;\beta ,\beta )}{P(f;0,\beta )}=\frac{\mathrm{e}^{-N\beta f}}{%
2^{N}\left( \cosh \beta \right) ^{N_{B}}}.  \label{FT4}
\end{equation}%
By taking the logarithm and average to make the form of the KL divergence,
we obtain two of inequalities as 
\begin{equation}
-\beta f^*(\beta ,\beta )\geq \log 2+d\log \left( \cosh \beta \right)\label{Ineq1}
\end{equation}%
and 
\begin{equation}
-\beta f^*(0,\beta )\leq \log 2+d\log \left( \cosh \beta \right),\label{Ineq2}
\end{equation}
where we defined the free energy in the thermodynamic limit as
\begin{equation}
f^*(\beta_p,\beta ) = \int df P(f;\beta_p ,\beta )f
\end{equation}
\subsection{Phase diagrams in spin glasses}
\begin{figure}[tb]
\begin{center}
\includegraphics[width=85mm]{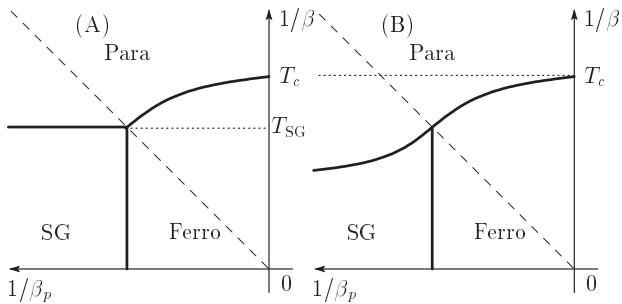}
\end{center}
\caption{{\protect\small Phase diagrams of spin glasses. (A) The typical phase
diagram of several mean-field models and that of Mattis model (then $T_{c}=T_{\mathrm{SG}}$). (B) typical instance of the finite-dimensional $\pm J$ Ising model. 
The vertical axis denotes the temperature and the horizontal one expresses the density of the quenched randomness. 
The dashed line depicts the Nishimori line ($\protect\beta _{p}=\protect\beta $). 
The solid lines separate the representative phases: the ferromagnetic (Ferro), paramagnetic (Para) and spin glass (SG) ones.
In (B), below the dotted line, the Griffiths paramagnetic phase is expected to be laid.}}
\label{fig1}
\end{figure}

The common quantity on the right-hand sides of Eqs. (\ref{Ineq1}) and (\ref{Ineq2}) is equal to the annealed free energy given in the symmetric distribution as $-\beta f_{a}(0,\beta )=\log [Z(\beta ;\{\tau_{ij}\})]_{\beta _{p}=0}/N$. 
Thus the inequalities (\ref{Ineq1}) and (\ref{Ineq2}) read 
\begin{equation}
f^*(\beta ,\beta )\leq f_{a}(0,\beta )
\end{equation}%
and 
\begin{equation}
f^*(0,\beta )\geq f_{a}(0,\beta ).
\end{equation}%
From the fluctuation-theorem type relation (\ref{FT4}), the violation of the
equality relates to the KL divergence as 
\begin{equation}
f^*(\beta ,\beta )-f_{a}(0,\beta )=-\frac{1}{\beta }D_{\mathrm{KL}}(P(f;\beta
,\beta )|P(f;0,\beta ))
\end{equation}
and 
\begin{equation}
f^*(0,\beta )-f_{a}(0,\beta )=\frac{1}{\beta }D_{\mathrm{KL}}(P(f;0,\beta
)|P(f;\beta ,\beta )).
\end{equation}
When $f^*(0,\beta )=f_{a}(0,\beta )$, we immediately find $D_{\mathrm{KL}}(P(f;\beta ,\beta )|P(f;0,\beta ))=D_{\mathrm{KL}}(P(f;0,\beta) |P(f;\beta,\beta ))=0$. 
Thus we conclude that $f^*(\beta ,\beta )=f^*(0,\beta )$. 
This relation ensures that the critical point on the Nishimori line is located at the same temperature in the symmetry distribution, when $f^*(0,\beta)=f_{a}(0,\beta )$ as often seen in the paramagnetic solutions for the mean-field spin glass models and the free energy of the Mattis model \cite{Mattis1976,Ozeki1993} as in the case of (A) in Fig. \ref{fig1}. 
These models show the parallel phase boundary to $\beta _{p}$-axis from the Nishimori line to the region in the symmetric distribution. 

In addition, since $f_{a}(0,\beta )$ is a trivial function, the non-analytical point of the free energy $f^*$ is identified as that of the KL divergence. 
If two of the KL divergences $D_{\mathrm{KL}}(P(f;\beta ,\beta )|P(f;0,\beta ))$ and $D_{\mathrm{KL}}(P(f;0,\beta )|P(f;\beta,\beta ))$ have the non-analytical points at the same temperature, the phase boundary can be parallel to $\beta _{p}$-axis from the Nishimori line to the region in the symmetric distribution. 
Notice that not necessarily it means two of the KL divergence coincide with each other. 
The case is expected to be the Griffiths singularity, which is considered to be located at the same temperature $T_{c}$ for any $\beta _{p}$ as the ferromagnetic transition point without the quenched randomness \cite{Griffiths1969,Randeria1985,Matsuda2008} as in the case of (B) in Fig. \ref{fig1}. 
In this sense, the Griffiths singularity might be specified as the appearance of the simultaneous non-analytical point of two symmetric KL divergences of the free-energy distribution functions. 
It would push up the understanding of the Griffiths singularity with the aid of the information geometry \cite{Amari2000} through the above consideration.

\section{Conclusion}

We analyzed the distribution function of the performed work for the spin glass with the gauge symmetry. 
The gauge symmetry revealed the existence of the symmetry in the rate function of the performed work as well as the free energy depending on each realization of the quenched randomness. 
As a result we obtained several relations in the same form as the fluctuation theorem. 
In order to analyze spin glass system, we usually use the replica method, which deals with the highly-correlated multiple system of the original model. 
Although a part of our results overlapped with the known properties via the replica method and gauge transformation as we recovered, our analyses was directly performed on the distribution function without replica method and related the averaged quantities as the performed work and free energy to the KL divergence. 
In this sense we believe that our analyses should be valuable to provide a different perspective to understand the nonequilibrium and critical behaviors for spin glasses.

\begin{acknowledgments}
The author thanks the fruitful discussions with Yuki Sughiyama, Tomoyuki
Obuchi, Koji Hukushima, Jun-ichi Inoue, and Hidetoshi Nishimori. This work
was supported by MEXT in Japan, Grant-in-Aid for Young Scientists (B)
No.24740263.
\end{acknowledgments}

\appendix
\section{Crook's fluctuation theorem}\label{AP1}
We here demonstrate the Crooks's fluctuation theorem \cite{Crooks1998} of the performed work employing derivation by use of the large-deviation principle as done for the long-time behaviors by the Lebowitz and Spohn \cite{Lebowitz1999}. 
Instead of the long time observation ($T\gg 1$), we consider the large number of components ($N \gg 1$) in the present study. 

We assume that the detailed balance condition is satisfied as 
\begin{equation}
\frac{P_t(\mathbf{S}_{t+1}|\mathbf{S}_{t},\{\tau_{ij}\})}{P_t(\mathbf{S}_{t}|%
\mathbf{S}_{t+1},\{\tau_{ij}\})} = \frac{P_{\mathrm{eq.}}^{t}(\mathbf{S}%
_{t+1}|\{\tau_{ij}\})}{P_{\mathrm{eq.}}^{t}(\mathbf{S}_{t}|\{\tau_{ij}\})}.
\end{equation}
Then the product over all steps can be expressed by 
\begin{eqnarray}
\prod_{t=0}^{T-1}\frac{P_t(\mathbf{S}_{t+1}|\mathbf{S}_{t},\{\tau_{ij}\})}{%
P_t(\mathbf{S}_{t}|\mathbf{S}_{t+1},\{\tau_{ij}\})} = \prod_{t=0}^{T-1} 
\frac{P_{\mathrm{eq.}}^{t}(\mathbf{S}_{t+1}|\{\tau_{ij}\})}{P_{\mathrm{eq.}%
}^{t}(\mathbf{S}_{t}|\{\tau_{ij}\})}  \notag \\
= \mathrm{e}^{- \sum_{t=0}^{T-1}\delta X(\mathbf{S}_{t+1},\mathbf{S}%
_{t}|\{\tau_{ij}\})},
\end{eqnarray}
where we defined the discretized pseudo heat as $\delta X_t(\mathbf{S}_{t+1},%
\mathbf{S}_{t}|\{\tau_{ij}\}) = \beta_t H(\mathbf{S}_{t+1}|\{\tau_{ij}\}) -
\beta_tH(\mathbf{S}_{t}|\{\tau_{ij}\})$. The (pseudo) heat is given by 
\begin{equation}
X(\{\mathbf{S}_{t}\},\{\tau_{ij}\}; 0 \to T) = \sum_{t=0}^{T-1} \delta X_t(\mathbf{S}_{t+1},%
\mathbf{S}_{t}|\{\tau_{ij}\}).
\end{equation}
Then we confirm the first law of thermodynamics as 
\begin{eqnarray}
Y(\{\mathbf{S}_{t}\},\{\tau_{ij}\}; 0 \to T) + X(\{\mathbf{S}_{t}\},\{\tau_{ij}\}; 0 \to T)  \notag \\
= \beta_T H({\bf S}_{T}|\{\tau_{ij}\}) - \beta_0 H({\bf S}_{0}|\{\tau_{ij}\}).
\end{eqnarray}
Therefore we reach a relation between the original process and its inverse one starting from the equilibrium states as 
\begin{eqnarray}
&&\prod_{t=0}^{T-1}P_t(\mathbf{S}_{t+1}|\mathbf{S}_{t},\{\tau_{ij}\})\mathrm{%
e}^{-Y(\{\mathbf{S}_{t}\},\{\tau_{ij}\}; 0 \to T)}  \notag \\
&&= \prod_{t=0}^{T-1}P_t(\mathbf{S}_{t}|\mathbf{S}_{t+1},\{\tau_{ij}\})%
\mathrm{e}^{-\beta_T H({\bf S}_{T}|\{\tau_{ij}\}) + \beta_0 H({\bf S}_{0}|\{\tau_{ij}\})}.  \notag \\
\end{eqnarray}
As a result, we find the following relation as
\begin{eqnarray}
&&\prod_{t=0}^{T-1}P_t(\mathbf{S}_{t+1}|\mathbf{S}_{t},\{\tau_{ij}\}) P_{%
\mathrm{eq.}}^{0}(\mathbf{S}_{0}|\{\tau_{ij}\})  \notag \\
&&= \frac{Z(\beta_T;\{\tau_{ij}\})}{Z(\beta_0;\{\tau_{ij}\})}  \notag \\
& & \quad \times \prod_{t=0}^{T-1}P_t(\mathbf{S}_{t}|\mathbf{S}%
_{t+1},\{\tau_{ij}\})P_{\mathrm{eq.}}^{T}(\mathbf{S}_{T}|\{\tau_{ij}\}) 
\mathrm{e}^{Ny}.  \notag \\
\end{eqnarray}
This relation yields Eq. (\ref{FT1}).

The rate function is given by the Legendre transformation of the generating
function as 
\begin{equation}
I(y; \{\tau_{ij}\}, 0 \to T) = \sup_r \left\{ r y - \Psi_y(r;
\{\tau_{ij}\},0 \to T) \right\}.
\end{equation}
On the other hand, the rate function for the inverse process is also defined
as 
\begin{equation}
I(y; \{\tau_{ij}\}, T \to 0) = \sup_r \left\{ r y - \Psi_y(r; \{\tau_{ij}\},
T \to 0) \right\}.
\end{equation}
By use of the relation (\ref{FT1}), we find 
\begin{eqnarray}
&&I(y; \{\tau_{ij}\}, 0 \to T)  \notag \\
&&\quad = \sup_r \left\{ (r+1)y - \Psi_y(-(r+1); \{\tau_{ij}\}, T \to 0)
\right\}  \notag \\
&&\qquad - y - \frac{1}{N} ( \log Z (\beta_T;\{\tau_{ij}\})- \log Z
(\beta_0;\{\tau_{ij}\}))  \notag \\
&&\quad = I(-y;\{\tau_{ij}\}, T \to 0)  \notag \\
&&\quad - y - \frac{1}{N} ( \log Z(\beta_T;\{\tau_{ij}\}) - \log
Z(\beta_0;\{\tau_{ij}\}) ).
\end{eqnarray}
This symmetry of the rate functions yields the well-known fluctuation
theorem for the distribution functions of the performed work as 
\begin{equation}
\frac{P(y;\{\tau_{ij}\},0 \to T)}{P(-y; \{\tau_{ij}\}, T \to 0)}=
\exp\left\{ N(y - \Delta_{0 \to T}(\beta f))\right\}.
\end{equation}
This is the Crook's fluctuation theorem \cite{Crooks1998}.
Notice that this derivation of the fluctuation theorem through the analysis of the generating function is trivial but has not been seen as far as the best of our knowledge.

\section{Gauge transformation}\label{AP2}
We demonstrate the manipulation of the gauge transformation to obtain Eq. (\ref{AGT1}) from Eq. (\ref{BGT1}).
The quantity in the second line of Eq. (\ref{BGT1}) can be written as
\begin{eqnarray}
&& \left[\mathrm{e}^{N \Psi_y(r;\{\tau_{ij}\},0 \to T)}\right]_{\beta_p} 
\notag \\
&& = \sum_{\{\tau_{ij}\}}\frac{\prod_{\langle ij \rangle}{\rm e}^{\beta_p\tau_{ij}}}{(2\cosh \beta_p)^{N_B}} \mathrm{e}^{N \Psi_y(r;\{\tau_{ij}\},0 \to T)}.
\end{eqnarray}
The gauge transformation yields
\begin{eqnarray}
&& \left[\mathrm{e}^{N \Psi_y(r;\{\tau_{ij}\},0 \to T)}\right]_{\beta_p} 
\notag \\
&& = \sum_{\{\tau_{ij}\}}\frac{\prod_{\langle ij \rangle}{\rm e}^{\beta_p\tau_{ij}\sigma_i\sigma_j}}{(2\cosh \beta_p)^{N_B}} \mathrm{e}^{N \Psi_y(r;\{\tau_{ij}\},0 \to T)}.
\end{eqnarray}
Thus we sum over all possible configurations of $\{\sigma_i\}$ and obtain
\begin{eqnarray}
&& 2^N \left[\mathrm{e}^{N \Psi_y(r;\{\tau_{ij}\},0 \to T)}\right]_{\beta_p} 
\notag \\
&& = \sum_{\{\tau_{ij}\}}\frac{Z(\beta_p;\{\tau_{ij}\})}{(2\cosh \beta_p)^{N_B}} \mathrm{e}^{N \Psi_y(r;\{\tau_{ij}\},0 \to T)}.
\end{eqnarray}
Similarly, we can evaluate the quantity in the third line of Eq. (\ref{BGT1}) as
\begin{eqnarray}
&& 2^N \left[\mathrm{e}^{N \Psi_y(r;\{\tau_{ij}\},0 \to T)}\right]_{\beta_p} 
\notag \\
&& = \sum_{\{\tau_{ij}\}}\frac{Z(\beta_T;\{\tau_{ij}\})}{Z(\beta_0;\{\tau_{ij}\})} \frac{Z(\beta_p;\{\tau_{ij}\})}{(2\cosh \beta_p)^{N_B}} \mathrm{e}^{N \Psi_y(-(r+1);\{\tau_{ij}\}, T \to 0)}. \notag \\
\end{eqnarray}
Therefore we reproduce Eq. (\ref{AGT1}).

In addition, the quantity in the second line of Eq. (\ref{BGT2}) is
\begin{eqnarray}
&& \left[\mathrm{e}^{N \Psi_y(r;\{\tau_{ij}\},T \to 0)}\right]_{\beta_p} 
\notag \\
&& = \sum_{\{\tau_{ij}\}}\frac{\prod_{\langle ij \rangle}{\rm e}^{\beta_p\tau_{ij}}}{(2\cosh \beta_p)^{N_B}} \mathrm{e}^{N \Psi_y(r;\{\tau_{ij}\},T \to 0)}.
\end{eqnarray}
The gauge transformation yields
\begin{eqnarray}
&& \left[\mathrm{e}^{N \Psi_y(r;\{\tau_{ij}\},T \to 0)}\right]_{\beta_p} 
\notag \\
&& = \sum_{\{\tau_{ij}\}}\frac{\prod_{\langle ij \rangle}{\rm e}^{\beta_p\tau_{ij}\sigma_i\sigma_j}}{(2\cosh \beta_p)^{N_B}} \mathrm{e}^{N \Psi_y(r;\{\tau_{ij}\},T \to 0)}.
\end{eqnarray}
The summation over all possible configurations of $\{\sigma_i\}$ yields
\begin{eqnarray}
&& 2^N \left[\mathrm{e}^{N \Psi_y(r;\{\tau_{ij}\},T \to 0)}\right]_{\beta_p} 
\notag \\
&& = \sum_{\{\tau_{ij}\}}\frac{Z(\beta_p;\{\tau_{ij}\})}{(2\cosh \beta_p)^{N_B}} \mathrm{e}^{N \Psi_y(r;\{\tau_{ij}\},T \to 0)}.
\end{eqnarray}
The same analysis can be applied to the third line of Eq. (\ref{BGT2}) as
\begin{eqnarray}
&& 2^N \left[\mathrm{e}^{N \Psi_y(r;\{\tau_{ij}\},T \to 0)}\right]_{\beta_p} 
\notag \\
&& = \sum_{\{\tau_{ij}\}}\frac{Z(\beta_0;\{\tau_{ij}\})}{Z(\beta_T;\{\tau_{ij}\})} \frac{Z(\beta_p;\{\tau_{ij}\})}{(2\cosh \beta_p)^{N_B}} \mathrm{e}^{N \Psi_y(-(r+1);\{\tau_{ij}\}, 0 \to T)}. \notag \\
\end{eqnarray}
When $\beta_p = \beta_T$, we find Eq. (\ref{Eq2}).

\bibliography{FTonNL_ver2}

\begin{thebibliography}{26}
\expandafter\ifx\csname natexlab\endcsname\relax\def\natexlab#1{#1}\fi
\expandafter\ifx\csname bibnamefont\endcsname\relax
  \def\bibnamefont#1{#1}\fi
\expandafter\ifx\csname bibfnamefont\endcsname\relax
  \def\bibfnamefont#1{#1}\fi
\expandafter\ifx\csname citenamefont\endcsname\relax
  \def\citenamefont#1{#1}\fi
\expandafter\ifx\csname url\endcsname\relax
  \def\url#1{\texttt{#1}}\fi
\expandafter\ifx\csname urlprefix\endcsname\relax\def\urlprefix{URL }\fi
\providecommand{\bibinfo}[2]{#2}
\providecommand{\eprint}[2][]{\url{#2}}

\bibitem[{\citenamefont{Crooks}(1998)}]{Crooks1998}
\bibinfo{author}{\bibfnamefont{G.~E.} \bibnamefont{Crooks}},
  \bibinfo{journal}{J. Stat. Phys.} \textbf{\bibinfo{volume}{90}},
  \bibinfo{pages}{1481} (\bibinfo{year}{1998}).

\bibitem[{\citenamefont{Evans et~al.}(1993)\citenamefont{Evans, Cohen, and
  Morriss}}]{Evans1993}
\bibinfo{author}{\bibfnamefont{D.~J.} \bibnamefont{Evans}},
  \bibinfo{author}{\bibfnamefont{E.~G.~D.} \bibnamefont{Cohen}},
  \bibnamefont{and} \bibinfo{author}{\bibfnamefont{G.~P.}
  \bibnamefont{Morriss}}, \bibinfo{journal}{Phys. Rev. Lett.}
  \textbf{\bibinfo{volume}{71}}, \bibinfo{pages}{2401} (\bibinfo{year}{1993}),
  \urlprefix\url{http://link.aps.org/doi/10.1103/PhysRevLett.71.2401}.

\bibitem[{\citenamefont{Evans and Searles}(1994)}]{Evans1994}
\bibinfo{author}{\bibfnamefont{D.~J.} \bibnamefont{Evans}} \bibnamefont{and}
  \bibinfo{author}{\bibfnamefont{D.~J.} \bibnamefont{Searles}},
  \bibinfo{journal}{Phys. Rev. E} \textbf{\bibinfo{volume}{50}},
  \bibinfo{pages}{1645} (\bibinfo{year}{1994}),
  \urlprefix\url{http://link.aps.org/doi/10.1103/PhysRevE.50.1645}.

\bibitem[{\citenamefont{Gallavotti and
  Cohen}(1995{\natexlab{a}})}]{Gallavotti1995b}
\bibinfo{author}{\bibfnamefont{G.}~\bibnamefont{Gallavotti}} \bibnamefont{and}
  \bibinfo{author}{\bibfnamefont{E.}~\bibnamefont{Cohen}},
  \bibinfo{journal}{Journal of Statistical Physics}
  \textbf{\bibinfo{volume}{80}}, \bibinfo{pages}{931}
  (\bibinfo{year}{1995}{\natexlab{a}}), ISSN \bibinfo{issn}{0022-4715},
  \bibinfo{note}{10.1007/BF02179860},
  \urlprefix\url{http://dx.doi.org/10.1007/BF02179860}.

\bibitem[{\citenamefont{Gallavotti and
  Cohen}(1995{\natexlab{b}})}]{Gallavotti1995a}
\bibinfo{author}{\bibfnamefont{G.}~\bibnamefont{Gallavotti}} \bibnamefont{and}
  \bibinfo{author}{\bibfnamefont{E.~G.~D.} \bibnamefont{Cohen}},
  \bibinfo{journal}{Phys. Rev. Lett.} \textbf{\bibinfo{volume}{74}},
  \bibinfo{pages}{2694} (\bibinfo{year}{1995}{\natexlab{b}}),
  \urlprefix\url{http://link.aps.org/doi/10.1103/PhysRevLett.74.2694}.

\bibitem[{\citenamefont{Kurchan}(1998)}]{Kurchan1998}
\bibinfo{author}{\bibfnamefont{J.}~\bibnamefont{Kurchan}},
  \bibinfo{journal}{Journal of Physics A: Mathematical and General}
  \textbf{\bibinfo{volume}{31}}, \bibinfo{pages}{3719} (\bibinfo{year}{1998}),
  \urlprefix\url{http://stacks.iop.org/0305-4470/31/i=16/a=003}.

\bibitem[{\citenamefont{Lebowitz and Spohn}(1999)}]{Lebowitz1999}
\bibinfo{author}{\bibfnamefont{J.~L.} \bibnamefont{Lebowitz}} \bibnamefont{and}
  \bibinfo{author}{\bibfnamefont{H.}~\bibnamefont{Spohn}},
  \bibinfo{journal}{Journal of Statistical Physics}
  \textbf{\bibinfo{volume}{95}}, \bibinfo{pages}{333} (\bibinfo{year}{1999}),
  ISSN \bibinfo{issn}{0022-4715}, \bibinfo{note}{10.1023/A:1004589714161},
  \urlprefix\url{http://dx.doi.org/10.1023/A:1004589714161}.

\bibitem[{\citenamefont{Maes}(1999)}]{Maes1999}
\bibinfo{author}{\bibfnamefont{C.}~\bibnamefont{Maes}},
  \bibinfo{journal}{Journal of Statistical Physics}
  \textbf{\bibinfo{volume}{95}}, \bibinfo{pages}{367} (\bibinfo{year}{1999}),
  ISSN \bibinfo{issn}{0022-4715}, \bibinfo{note}{10.1023/A:1004541830999},
  \urlprefix\url{http://dx.doi.org/10.1023/A:1004541830999}.

\bibitem[{\citenamefont{Gaspard}(2012)}]{Gaspard2012}
\bibinfo{author}{\bibfnamefont{P.}~\bibnamefont{Gaspard}},
  \bibinfo{journal}{Journal of Statistical Mechanics: Theory and Experiment}
  \textbf{\bibinfo{volume}{2012}}, \bibinfo{pages}{P08021}
  (\bibinfo{year}{2012}),
  \urlprefix\url{http://stacks.iop.org/1742-5468/2012/i=08/a=P08021}.

\bibitem[{\citenamefont{Jarzynski}(1997{\natexlab{a}})}]{Jarzynski1997b}
\bibinfo{author}{\bibfnamefont{C.}~\bibnamefont{Jarzynski}},
  \bibinfo{journal}{Phys. Rev. E} \textbf{\bibinfo{volume}{56}},
  \bibinfo{pages}{5018} (\bibinfo{year}{1997}{\natexlab{a}}),
  \urlprefix\url{http://link.aps.org/doi/10.1103/PhysRevE.56.5018}.

\bibitem[{\citenamefont{Jarzynski}(1997{\natexlab{b}})}]{Jarzynski1997a}
\bibinfo{author}{\bibfnamefont{C.}~\bibnamefont{Jarzynski}},
  \bibinfo{journal}{Phys. Rev. Lett.} \textbf{\bibinfo{volume}{78}},
  \bibinfo{pages}{2690} (\bibinfo{year}{1997}{\natexlab{b}}),
  \urlprefix\url{http://link.aps.org/doi/10.1103/PhysRevLett.78.2690}.

\bibitem[{\citenamefont{Ohzeki et~al.}(2011)\citenamefont{Ohzeki, Katsuda, and
  Nishimori}}]{Ohzeki2011b}
\bibinfo{author}{\bibfnamefont{M.}~\bibnamefont{Ohzeki}},
  \bibinfo{author}{\bibfnamefont{H.}~\bibnamefont{Katsuda}}, \bibnamefont{and}
  \bibinfo{author}{\bibfnamefont{H.}~\bibnamefont{Nishimori}},
  \bibinfo{journal}{Journal of the Physical Society of Japan}
  \textbf{\bibinfo{volume}{80}}, \bibinfo{pages}{084002}
  (\bibinfo{year}{2011}),
  \urlprefix\url{http://jpsj.ipap.jp/link?JPSJ/80/084002/}.

\bibitem[{\citenamefont{Ohzeki and Nishimori}(2010)}]{Ohzeki2010b}
\bibinfo{author}{\bibfnamefont{M.}~\bibnamefont{Ohzeki}} \bibnamefont{and}
  \bibinfo{author}{\bibfnamefont{H.}~\bibnamefont{Nishimori}},
  \bibinfo{journal}{J. Phys. Soc. Jpn.} \textbf{\bibinfo{volume}{79}},
  \bibinfo{pages}{084003} (\bibinfo{year}{2010}),
  \urlprefix\url{http://jpsj.ipap.jp/link?JPSJ/79/084003/}.

\bibitem[{\citenamefont{Nishimori}(1981)}]{Nishimori1981}
\bibinfo{author}{\bibfnamefont{H.}~\bibnamefont{Nishimori}},
  \bibinfo{journal}{Progress of Theoretical Physics}
  \textbf{\bibinfo{volume}{66}}, \bibinfo{pages}{1169} (\bibinfo{year}{1981}),
  \urlprefix\url{http://ptp.ipap.jp/link?PTP/66/1169/}.

\bibitem[{\citenamefont{Nishimori}(2001)}]{HNbook}
\bibinfo{author}{\bibfnamefont{H.}~\bibnamefont{Nishimori}},
  \emph{\bibinfo{title}{Statistical physics of spin glasses and information
  processing : an introduction}} (\bibinfo{publisher}{Oxford University Press},
  \bibinfo{address}{Oxford New York}, \bibinfo{year}{2001}), ISBN
  \bibinfo{isbn}{0198509413}.

\bibitem[{\citenamefont{Kirkpatrick et~al.}(1983)\citenamefont{Kirkpatrick,
  Gelatt, and Vecchi}}]{Kirkpatrick1983}
\bibinfo{author}{\bibfnamefont{S.}~\bibnamefont{Kirkpatrick}},
  \bibinfo{author}{\bibfnamefont{C.~D.} \bibnamefont{Gelatt}},
  \bibnamefont{and} \bibinfo{author}{\bibfnamefont{M.~P.}
  \bibnamefont{Vecchi}}, \bibinfo{journal}{Science}
  \textbf{\bibinfo{volume}{220}}, \bibinfo{pages}{671} (\bibinfo{year}{1983}),
  \eprint{http://www.sciencemag.org/content/220/4598/671.full.pdf},
  \urlprefix\url{http://www.sciencemag.org/content/220/4598/671.abstract}.

\bibitem[{\citenamefont{Ozeki}(1995)}]{Ozeki1995}
\bibinfo{author}{\bibfnamefont{Y.}~\bibnamefont{Ozeki}},
  \bibinfo{journal}{Journal of Physics A: Mathematical and General}
  \textbf{\bibinfo{volume}{28}}, \bibinfo{pages}{3645} (\bibinfo{year}{1995}),
  \urlprefix\url{http://stacks.iop.org/0305-4470/28/i=13/a=010}.

\bibitem[{\citenamefont{Ohzeki}(2010)}]{Ohzeki2010a}
\bibinfo{author}{\bibfnamefont{M.}~\bibnamefont{Ohzeki}},
  \bibinfo{journal}{Phys. Rev. Lett.} \textbf{\bibinfo{volume}{105}},
  \bibinfo{pages}{050401} (\bibinfo{year}{2010}),
  \urlprefix\url{http://link.aps.org/doi/10.1103/PhysRevLett.105.050401}.

\bibitem[{\citenamefont{{Georges, A.} et~al.}(1987)\citenamefont{{Georges, A.},
  {Hansel, D.}, {Le Doussal, P.}, and {Maillard, J.M.}}}]{Georges1987}
\bibinfo{author}{\bibnamefont{{Georges, A.}}},
  \bibinfo{author}{\bibnamefont{{Hansel, D.}}},
  \bibinfo{author}{\bibnamefont{{Le Doussal, P.}}}, \bibnamefont{and}
  \bibinfo{author}{\bibnamefont{{Maillard, J.M.}}}, \bibinfo{journal}{J. Phys.
  France} \textbf{\bibinfo{volume}{48}}, \bibinfo{pages}{1}
  (\bibinfo{year}{1987}),
  \urlprefix\url{http://dx.doi.org/10.1051/jphys:019870048010100}.

\bibitem[{\citenamefont{Ohzeki and Nishimori}(2009)}]{Ohzeki2009b}
\bibinfo{author}{\bibfnamefont{M.}~\bibnamefont{Ohzeki}} \bibnamefont{and}
  \bibinfo{author}{\bibfnamefont{H.}~\bibnamefont{Nishimori}},
  \bibinfo{journal}{Journal of Physics A: Mathematical and Theoretical}
  \textbf{\bibinfo{volume}{42}}, \bibinfo{pages}{332001}
  (\bibinfo{year}{2009}),
  \urlprefix\url{http://stacks.iop.org/1751-8121/42/i=33/a=332001}.

\bibitem[{\citenamefont{Mattis}(1976)}]{Mattis1976}
\bibinfo{author}{\bibfnamefont{D.}~\bibnamefont{Mattis}},
  \bibinfo{journal}{Physics Letters A} \textbf{\bibinfo{volume}{56}},
  \bibinfo{pages}{421 } (\bibinfo{year}{1976}), ISSN \bibinfo{issn}{0375-9601},
  \urlprefix\url{http://www.sciencedirect.com/science/article/pii/037596017690%
3960}.

\bibitem[{\citenamefont{Ozeki}(1993)}]{Ozeki1993}
\bibinfo{author}{\bibfnamefont{Y.}~\bibnamefont{Ozeki}},
  \bibinfo{journal}{Journal of Statistical Physics}
  \textbf{\bibinfo{volume}{71}}, \bibinfo{pages}{759} (\bibinfo{year}{1993}),
  ISSN \bibinfo{issn}{0022-4715}, \bibinfo{note}{10.1007/BF01058446},
  \urlprefix\url{http://dx.doi.org/10.1007/BF01058446}.

\bibitem[{\citenamefont{Griffiths}(1969)}]{Griffiths1969}
\bibinfo{author}{\bibfnamefont{R.~B.} \bibnamefont{Griffiths}},
  \bibinfo{journal}{Phys. Rev. Lett.} \textbf{\bibinfo{volume}{23}},
  \bibinfo{pages}{17} (\bibinfo{year}{1969}),
  \urlprefix\url{http://link.aps.org/doi/10.1103/PhysRevLett.23.17}.

\bibitem[{\citenamefont{Matsuda et~al.}(2008)\citenamefont{Matsuda, Nishimori,
  and Hukushima}}]{Matsuda2008}
\bibinfo{author}{\bibfnamefont{Y.}~\bibnamefont{Matsuda}},
  \bibinfo{author}{\bibfnamefont{H.}~\bibnamefont{Nishimori}},
  \bibnamefont{and}
  \bibinfo{author}{\bibfnamefont{K.}~\bibnamefont{Hukushima}},
  \bibinfo{journal}{Journal of Physics A: Mathematical and Theoretical}
  \textbf{\bibinfo{volume}{41}}, \bibinfo{pages}{324012}
  (\bibinfo{year}{2008}),
  \urlprefix\url{http://stacks.iop.org/1751-8121/41/i=32/a=324012}.

\bibitem[{\citenamefont{Randeria et~al.}(1985)\citenamefont{Randeria, Sethna,
  and Palmer}}]{Randeria1985}
\bibinfo{author}{\bibfnamefont{M.}~\bibnamefont{Randeria}},
  \bibinfo{author}{\bibfnamefont{J.~P.} \bibnamefont{Sethna}},
  \bibnamefont{and} \bibinfo{author}{\bibfnamefont{R.~G.}
  \bibnamefont{Palmer}}, \bibinfo{journal}{Phys. Rev. Lett.}
  \textbf{\bibinfo{volume}{54}}, \bibinfo{pages}{1321} (\bibinfo{year}{1985}),
  \urlprefix\url{http://link.aps.org/doi/10.1103/PhysRevLett.54.1321}.

\bibitem[{\citenamefont{Amari}(2000)}]{Amari2000}
\bibinfo{author}{\bibfnamefont{S.}~\bibnamefont{Amari}},
  \emph{\bibinfo{title}{Methods of information geometry}}
  (\bibinfo{publisher}{American Mathematical Society},
  \bibinfo{address}{Providence, RI}, \bibinfo{year}{2000}), ISBN
  \bibinfo{isbn}{9780821805312}.

\end{thebibliography}

\end{document}